\documentclass{article}

\usepackage{fullpage}
\usepackage[latin1]{inputenc}
\usepackage[T1]{fontenc}
\usepackage[10pt]{type1ec}
\usepackage[polutonikogreek,english]{babel}
\usepackage{amsfonts}
\usepackage{amssymb}
\usepackage{amsmath}
\usepackage{amsthm}
\usepackage{graphicx,xcolor}
\usepackage{afterpage}
\usepackage{bbm}
\usepackage{latexsym}
\definecolor{sap}{RGB}{129,36,51}

\newtheorem*{teorema}{Theorem}
\newtheorem*{Hyp}{Hypotheses on the interaction}

\def\be{\begin{equation}}
\def\ee{\end{equation}}
\def\bea{\begin{eqnarray}}
\def\eea{\end{eqnarray}}
\def\ni{\noindent}
\def\nn{\nonumber}

\def\sx{\sigma^x}
\def\sy{\sigma^y}
\def\sz{\sigma^z}

\def\a{\alpha}

\def\d{\delta}
\def\g{\gamma}
\def\G{\Gamma}
\def\b{\beta}
\def\D{\Delta}

\def\t{\tau}
\def\dg{\dagger}

\def\w{\omega}
\def\wq{\omega_q}
\def\wp{\omega_p}
\def\Tr{\operatorname{Tr}}

\newcommand{\ie}{\textit{i.e. }}
\newcommand{\eg}{\textit{e.g. }}
\newcommand{\meanv}[1]{\left\langle#1\right\rangle}
\newcommand{\ket}[1]{\left|#1\right\rangle}

\newcommand{\nocontentsline}[3]{}
\newcommand{\tocless}[2]{\bgroup\let\addcontentsline=\nocontentsline#1{#2}\egroup}

\def\Hom{\operatorname{Hom}}
\def\Inhom{\operatorname{Inhom}}

\title{On the Dynamics of XY Spin Chains with Impurities}
\author{Giuseppe Genovese \footnote{email: giuseppe.genovese@math.uzh.ch}\\  \\  \textit{Insitut f\"ur Mathematik, Universit\"at Z\"urich}\\ 
\textit{Winterthurerstrasse 190, CH-8057 Z\"urich, CH}
}

\begin{document}

\maketitle

\begin{abstract}
\ni We provide a theoretical set up for studying the dynamics in quantum spin chain models with inhomogeneous two-body interaction. We frame in our formalism models that can be mapped into fermion systems with quadratic Hamiltonian, namely XY chains with transverse field. Local and global existence results of the dynamics are discussed.
\end{abstract}

\section{Introduction and Summary}

\ni In this paper we will treat time dependent Hamiltonians in many body quantum mechanics. Since the subject is intrinsically problematic, we will analyse one of the simplest significant class of models available, namely quasi-free Fermi particle systems in one dimension.

\ni Nowadays quantum mechanics is a common background to diverse branches of science, and time dependent formalism finds different attitudes in the community. However it arises quite naturally in statistical mechanics in the context of non equilibrium dynamics. 

\ni The philosophy of non autonomous systems is in few words that the action of the environment on the system is described in an effective theory by putting a time dependence in the Hamiltonian. This is fairly a coarse approximation, since it does not take into account the structure of the environment, nor the effect that the system has on it. Nonetheless this picture is very helpful for many practical purposes, and it deserves certainly to be thoroughly studied to improve our comprehension of non equilibrium phenomena.

\ni The XY spin chains constitute a very appropriate model to study the basic features of many body quantum systems, and indeed there is a sizeable literature on the topic (see for instance \cite{ABGM}, \cite{B2}, \cite{Bar}, \cite{BMD}, \cite{cal}, \cite{Cir}, \cite{Raf}, \cite{algebra}, \cite{Ig}, \cite{LMS}, \cite{perk}, \cite{ros}, \cite{Nat} and reference therein, albeit this list is far to be exhaustive). In particular the dynamical properties of such models are non trivial. For example the global transverse magnetisation of the XY model has been observed (see \cite{BMD} or \cite{Bar}) to approach a non equilibrium limit when a global transverse field is abruptly switched on. This kind of features has been the subject of an intensive study by the '70: we refer for reviews to \cite{Leb}\cite{Bar} and for more recent developments to \cite{Ig}\cite{ros} (and references therein). 

\ni More recently spin chains have gained increasing interest from different perspectives: \textit{e.g.} in the statistical mechanics of non equilibrium phenomena they provide a simple quantum model for studying the effect of reservoirs on an extensive system (in the set-up presented for instance in \cite{algebra} or \cite{Marcus}), along with the properties of the quantum phase transition associated to these models (as for instance in \cite{Cir}); from the quantum information side one dimensional spin systems are appealing test models for information transfer protocols, also because nowadays we have experimental realisations (with few atoms) of such systems \cite{Nat}.

\ni On the other hand impurities change drastically the scenario, both in equilibrium and in dynamics. They destroy many of the symmetries of the model, so rendering thermodynamics very difficult to analyse. At present a clear picture of the equilibrium features of spin chains with inhomogeneous couplings is actually unavailable, and, moreover, it appears to be very far to be achieved, at least from the viewpoint of rigorous statistical mechanics. In particular, it seems to be a priority to shed more light on the Griffith-McCoy smooth phase transition \cite{GMc} that is expected to be induced by the disorder \cite{Raf}. 

\ni Of course, since equilibrium properties of such a class of systems still conserve many aspects not investigated, it is even harder to study the dynamics. The main difficulty is the lack of informations on the structure of the spectrum of the Hamiltonian at a fixed time. But also in the simplest occurrence of one sole impurity (as examined in \cite{ABGM}), the presence of an isolated eigenvalue causes many technical difficulties. We will return on this point in the sequel. 

\ni The goal of the paper is to provide explicit equations for the dynamics of the XY chain in transverse field with inhomogeneous coupling. The dynamics is encoded in an apparently complicate system of integral equations. Since this model is a particular case of quadratic Hamiltonian in the Fermi operators, we will discuss the specific properties of the dynamics in relation to more general results. It turns out that, even though a local existence theorem arises quite naturally for such systems, the extension to global existence is in general much more delicate. This can be thought as a direct consequence of the lack of conservation laws in time dependent theories. 

\ni Our aim is therefore to englobe the variety of phenomena studied in the large literature existing on the topic in a unified formalism. Some similar attempts can be found in \cite{B2}, \cite{cal}, \cite{ros} and \cite{Ig} especially for quenches (which are included in our analysis). In the evolution equation we obtain, there are encoded many remarkable dynamical aspects, as time irreversibility, occurrence of memory effect and the consequent break of ergodicity. Albeit this features were much studied in the homogeneous setting, only a limited attention has been given to the systems with impurities. 

\ni However we stress that in this work we limit ourself to consider impurities as inhomogeneous perturbations of the XY chain. This provides quite a toy model to describe the variety of physical phenomena associated to defects (smoothing of the transition, Kondo effects... ), and so it captures only a part of the interesting features of many body disordered systems. However the study of such a class of systems has to be intended as a step toward the comprehension of more realistic ones. 

\vspace{0.5cm}

\ni The paper is organised as follows:

\ni In Section 2 we present the theoretical picture for the dynamics of systems ruled by a quadratic Hamiltonian in annihilation and creation operators for both bosons and fermions.

\ni In Section 3 we will be concerned with states and observables: in particular we will give the explicit form of the time dependent state of the system. 

\ni In Section 4 we introduce the standard basic concepts about quantum spin chains and the mapping into fermions.

\ni In Section 5 we will use results in Section 2 in the framework of quantum spin chains. We introduce suitable variables in order to capture the effect given by the presence of impurities and we present the system of integral equations associated to the dynamics.

\ni In Section 6 we will deal with specific cases, displaying explicitly the equations of dynamics for the anisotropic and isotropic XY model in an inhomogeneous transverse field, and we will discuss rather concisely the motion of a single impurity.

\ni Section 7 is left to further discussions and outlooks.

\ni Furthermore we attach two appendices: Appendix \ref{app-A} presents a succinct survey of the diagonalisation method of the XY Hamiltonian, after the seminal paper of Lieb, Schultz and Mattis \cite{LMS}; Appendix \ref{app:B} shows a simple (but interesting) situation in which the whole formalism can be applied, namely the quench of an impurity in the XX chain. 


\section{Quadratic Hamiltonians: Set Up and Dynamics}

\ni Our starting point is a system of $N$ quantum particles on the grid. We do not still specify the statistics of the particles: we describe them by creation and annihilation operators $c,c^\dg$, satisfying the canonical anti-commutation rules ($+$ sign) or the canonical commutation rules ($-$ sign), respectively if we are dealing with a system of particle obeying to Fermi-Dirac or Bose-Einstein statistics:
\be
\left[c_j,c_k\right]_{\pm}=0,\quad\left[c^\dg_j,c_k\right]_{\pm}=\d_{jk}.
\ee
We attach a quantum particle to each site, mathematically a Hilbert space $\mathcal{H}_i$, and the states space of the system is the Fock space $\mathcal{F}\equiv\bigoplus_N\bigotimes_{i=1}^N\mathcal{H}_i$. We are interested in general quadratic time dependent Hamiltonians:
\be\label{eq:H}
H_N(t)=\sum_{j,k\in\mathbb{Z}}J_{jk}(t)c^\dg_jc_k+\frac12K^*_{jk}(t)c^\dg_jc^\dg_k+\frac12K_{jk}(t)c_jc_k.
\ee

\begin{Hyp}
\ni We require: 
\begin{enumerate}
\item stability: $J_{jk}(t)\geq J_0>-\infty$;
\item $\ell_1$ finiteness: $\sup_k\sum_j|J_{jk}(t)|,\;\sup_j\sum_k|J_{jk}(t)|<\infty$ and $\sup_k\sum_j|K_{jk}(t)|,\;\sup_j\sum_k|K_{jk}(t)|<\infty$, uniformly in $t$;
\item boundedness and piecewise continuity in $t\in[t_0,T)$.
\end{enumerate}
\end{Hyp}

\ni The last hypothesis will allow us to use Fourier inverse transform also in $t$ (in the meaning of distributions).  Of course $J(t)$ is a Hermitian matrix and $K(t)$ is symmetric or anti-symmetric according to the statistics we are considering.

\ni These requirements make it possible to pass to Fourier transform: the Hamiltonian becomes

$$
H_N(t)=\frac{1}{N^2}\sum_{q,p\in\mathbb{Z}(N)}\a_{qp}(t)a^\dg_qa_p+\frac12\b_{qp}(t)a^\dg_q a_p^\dg+\frac12\b^*_{qp}(t)a_qa_p,
$$
where $\mathbb{Z}(N)$ denotes the set $\{(2\pi k)/N,\: k=1,\dots, N\}$ and
\be\label{eq:FT1}
\left\{
\begin{array}{rrr}
c_j&=&\frac{1}{N}\sum_{q\in\mathbb{Z}(N)}e^{-iqj}a_q\\
c^\dg_j&=&\frac{1}{N}\sum_{q\in\mathbb{Z}(N)}e^{iqj}a^\dg_q
\end{array}
\right.
;\qquad
\left\{
\begin{array}{rrr}
a_q&=&\sum_{j=1}^Ne^{iqj}c_j\\
a^\dg_q&=&\sum_{j}e^{-iqj}c^\dg_j
\end{array}
\right.
\ee
and
\be\label{eq:alfa-beta}
\a_{qp}(t)\equiv\sum_{j,k}J_{jk}e^{i(qj-pk)};\quad \b_{qp}(t)\equiv\sum_{j,k}K_{jk}e^{i(qj+pk)}.
\ee
\ni The fundamental feature of this Hamiltonian is that for $t'\neq t''$ in general $[H_N(t'),H_N(t'')]\neq0$. As it is well known, this condition implies a non trivial dynamics. We can regard the time evolution of our system as a dynamical system indexed by $q\in\mathbb{Z}(N)$ with the Heisenberg equations of motion
$$
d_t a_q(t)=[a_q(t),-iH(t)], \quad q\in\mathbb{Z}(N).
$$
These are the equations for the evolution semigroup. We will show that it is actually given by a one-parameter semigroup of Bogoliubov-Valatin transformations \cite{BV}, namely
\be\label{eq:B-V1par}
a_q(t)=\frac1N\sum_p A_{qp}(t)a_p+B_{qp}(t)a^\dg_p,
\ee
where $A_{qp}(t), B_{qp}(t)$ become $L_2$ function on $[0,2\pi]\times[0,2\pi]$ for $N\to\infty$ uniformly in $t$ and $A_{qp}(t_0)=\d_{qp}$ and $B_{qp}(t_0)=0$. We seek a solution satisfying
\be\label{eq:BVflux1}
id_t a_q(t)=\frac1N\sum_p i\dot A_{qp}(t)a_p+i\dot B_{qp}(t)a^\dg_p.
\ee
We want to evaluate the commutator in the Heisenberg equations and compare it with the last formula. This is easily done: we notice
\bea
\left[a_p, a^\dg_{q'}a_{p'}\right]_\mp&=&\d_{q'p}a_p'\nn\\
\left[a_p, a^{\dg}_{q'}a^{\dg}_{p'}\right]_\mp&=&\d_{q'p}a^{\dg}_{p'}\pm\d_{pp'}a^{\dg}_{q'}\nn\\
\left[a_p, a_{q'}a_{p'}\right]_\mp&=&0\nn\\
\left[a^\dg_p, a^\dg_{q'}a_{p'}\right]_\mp&=&\pm\d_{pp'}a^\dg_{q'}\nn\\
\left[a^\dg_p, a^\dg_{q'}a^\dg_{p'}\right]_\mp&=&0\nn\\
\left[a^\dg_p, a_{q'}a_{p'}\right]_\mp&=&\d_{pp'}a_{q'}\pm\d_{pq'}a_{p'},\nn
\eea
and hence
\bea
id_ta_q(t)&=&\frac{1}{N^2}\sum_{q'p'}a_{p'}\left(A_{qq'}(t)\a_{q'p'}(t)+\frac{1}{2}B_{qq'}(t)(\b^*_{q'p'}(t)\pm\b^*_{p'q'}(t))\right)\nn\\
&+&\frac{1}{N^2}\sum_{q'p'}a^\dg_{p'}\left(\pm B_{qq'}(t)\a_{q'p'}(t)+\frac{1}{2}A_{qq'}(t)(\b_{q'p'}(t)\pm\b_{p'q'}(t))\right).\nn
\eea
This leads us to the following linear system of first order ordinary differential equations
\be\label{eq:BVflux2}
\left\{
\begin{array}{rrr}
i\dot A_{qp}(t)&=&\frac1N\sum_{q'}A_{qq'}(t)\a_{q'p}(t)+B_{qq'}(t)\b^*_{q'p}(t)\\
i\dot B_{qp}(t)&=&\pm\frac1N\sum_{q'}B_{qq'}(t)\a_{q'p}(t)+A_{qq'}(t)\b_{q'p}(t).
\end{array}
\right.
\ee
So the system (\ref{eq:BVflux2}) represents the evolution equations for the one-parameter semigroup of Bogoliubov-Valatin transformations that gives the quantum dynamics. It is perhaps remarkable that this structure, quite natural for quadratic Hamiltonians on the grid (see also \cite{narn} and \cite{Cir}), emerges also in other scenarios, as shown for instance in the more involved problem of effective dynamics for mean field interacting particles \cite{ben}. 

\ni As long as $N$ remains finite, classical theorems on linear systems of first order ordinary differential equations assure existence and uniqueness of a global solution in the interval of definition $[t_0,T]$. This solves completely the problem of existence of dynamics at finite size. On the other hand, to find an explicit form for the time evolution is a more challenging task. A general approach in literature to cope with this kind of problems is by series expansions of the solution \cite{Magno}.

\ni Of course the thermodynamic limit $N\to\infty$ requires much more attention. In the limit we are led to consider the integro-differential system 
\be\label{eq:IntDiff}
\left\{
\begin{array}{lll}
i\dot A(q,p; t)&=&\int_{0}^{2\pi}dq' \left(A(q,q'; t)\a(q',p; t)+B(q,q'; t)\b^*(q',p; t)\right)\\
\pm i\dot B(q,p; t)&=&\int_{0}^{2\pi}dq'\left(B(q,q'; t)\a(q',p; t)+A(q,q'; t)\b(q',p; t)\right)\\
A(q,p; t_0)&=&\d(q-p)\\
B(q,p; t_0)&=&0,
\end{array}
\right.
\ee
where $\a(q',p; t), \b(q',p; t)$ are $C_{q,p}C_t([0,2\pi]^{\times2}\times[t_0,T])$ functions, obtained as natural limit of $\a_{q'p}(t), \b_{q'p}(t)$ (that is straightforward by our hypotheses). It is suggestive here to represent the above system as
\be
\frac{d}{dt} \mathcal{A} = \G(t)\mathcal{A}(t),
\ee
that stands for
\be
\frac{d}{dt}\left(
\begin{array}{r}
A(q,p; t)\\
B(q,p; t)
\end{array}
\right)=\int_{0}^{2\pi} dq' 
\left(
\begin{array}{rr}
-i\a(q',p; t)&-i\b^*(q',p; t)\\
\mp i\b(q',p; t)&\mp i\a(q',p; t)
\end{array}
\right)
\left(
\begin{array}{r}
A(q,q'; t)\\
B(q,q'; t)
\end{array}
\right).
\ee
Let us analyse at first the simplest case in which $\G(t)=\G(t_0)$ is constant in time. It is evidently a compact linear map from 
the space $L_{2}([0,2\pi]^{2})$ of square integrable functions on $[0,2\pi]\times [0,2\pi]$ into itself. The dynamics exists globally, and the semigroup is given by exponentiation. This is the starting point of the analysis performed in \cite{narn}.

\ni In dealing with time dependent generators of the dynamics, the easiest occurrence is that as long as $t$ runs in its interval of definition, $\G(t)$ remains bounded and the solution keeps lying into its domain. By our hypotheses this is straightforward to verify in every interval $[t',t'']\subset[t_0,T)$ such that the couplings are continuos. In this way $\G(t)$ is a strongly continuous operator from $L_{2}C_t([0,2\pi]^{2}\times[t',t''])$ into itself and one can derive local existence of the dynamics from general theorems (for detailed discussion on the topic see for instance \cite{pazy}\cite{Si}):
\begin{teorema}[Local Existence]
Consider a quadratic Hamiltonian as in (\ref{eq:H}), with summable and stable couplings and piece-wise continuous dependance on time in $t\in[t_0,T)$, according to Hypotheses 1, 2, 3. Then in each interval $[t',t'']\subset[t_0,T)$ in which the Hamiltonian is continuous in $t$, the quantum dynamics is defined as one-parameter continuously $t$-differentiable semigroup of Bogoliubov--Valatin transformations.
\end{teorema}

\ni What is more challenging of course is to extend the local existence result to a global one. As a preliminary observation, we see that $\G(t)$ represents a family of compact operators in $L_2([0,2\pi]^{\times2})$ for $t\in[t_0,T)$, uniformly bounded in $[t_0,T]$. Therefore the spectrum of $\G(t)$ is uniformly bounded in $t$: the Hille-Yoshida theorem allows us to conclude that $\G(t)$ is a family of generators of strongly continuous semigroups of contractions and there is a number $w$ such that the resolvent set of $\G(t)$ is contained in $[w,+\infty)$ uniformly in $t$. Another crucial property we have for free is that the domain of the generators is the entire space $L_2([0,2\pi]^{\times 2})$ independently of $t$. Unfortunately, since the family $\G(t)$ is not by assumptions strongly continuous, this is not enough to ensure existence of the dynamics for all times. Anyway we can recover a similar property by requiring that in each discontinuity point of $\G(t)$ in $[t_0,T)$ the left and right limit exists. In this way, we can establish a local existence theorem locally around each discontinuity, because each interval containing a discontinuity splits into two intervals in which the previous local theorem holds, and then we can paste together the contributions. We sketch an argument somewhat more explicit: let $t^*$ be a discontinuity point, we focus on the interval $[t^*-\d, t^*+\d]$ for an arbitrary $\d>0$, such that the unique discontinuity falls at time $t^*$. We write for every $t\in[t^*-\d, t^*+\d]$
\bea
\mathcal{A}(t)&=&\mathcal{A}(t^*-\d)+\int_{t^*-\d}^{t}dt' \G(t')\mathcal{A}(t')\nn\\
&=&\mathcal{A}(t^*-\d)+\int_{t^*-\d}^{T}dt' \chi([t^*-\d,t])\G(t')\mathcal{A}(t')\nn.
\eea
Now we can make discrete approximations of $\G(t)$ in $[t^*-\d, t^*+\d]$ (that is possible in virtue to the required regularity): for a given even integer $N>0$ we divide $[t^*-\d, t^*+\d]$ in N intervals of length $\t=\frac{2\d}{N}$ and we have a discretised version of the last formula:
$$
\mathcal{A}_h-\t\sum_{k=1}^{N} \G_{kh}\mathcal{A}_k=\mathcal{A}(t^*-\d),
$$
with $t=h\t$, $\mathcal{A}_k=\mathcal{A}(k\t)$ and $\G_{kh}=I(k\leq h)\G(k\t)$. This is a linear system of equations, where $\G$ is a diagonal matrix with only the first $h$ entries non zero; it has a unique solution provided that $\det(\mathbb{I}-\t\G)\neq0$. In our setting, due to the freedom in the choice of $\d$, it is not hard to verify this condition uniformly in $N$. Then one can take the limit $N\to\infty$ to have the existence of the solution for the flux.

\ni Thus we have local existence also in presence of (bounded) discontinuities and we can extend the previous theorem to all intervals $[t',t'']\subset[t_0,T)$. In order to recover the whole $[t_0,T]$ we need a compactification condition, namely that the limit $\lim_{t\to T^-} \G(t)$ exists (it is finite by hypothesis). In this way one obtains existence of the flux for all $t\in[t_0,T]$ by standard arguments.  

\ni We summarise our results in the following
\begin{teorema}[Global Existence]
Consider a quadratic Hamiltonian as in (\ref{eq:H}), with summable and stable couplings and piece-wise continuous dependance on time in $t\in[t_0,T)$, according to Hypotheses 1, 2, 3, with a countable number of discontinuity points $t^*_i$. Additionally we assume that the left and right limits of the Hamiltonian in each discontinuity point $t^*_i$ exist, and furthermore the existence of the left limit $t\to T^-$ as well. Then the dynamics is globally defined in $[t_0,T]$ as one-parameter semigroup of Bogoliubov-Valatin transformations.
\end{teorema}

\ni The main advantage in this context is that one never deals with unbounded generators. This permits us to profit from piece-wise compactness, and so we need to require just suitable conditions to glue the solutions. The very question of existence of the dynamics therefore relies in our opinion only in the asymptotic behaviour for $t\to T$. We will discuss this point later on with the aid of a specific example.


\section{Quadratic Fermion Hamiltonians: States}

\ni The derivation presented above is very general in its simplicity, not depending on the particular quantum statistics of the particles. However, being interested in quantum spin, hereafter we will deal exclusively with fermions.

\ni Of course, all the information for the temporal dependance of observables, and thus, the state of the system, is encoded into the equations (\ref{eq:BVflux2}) at finite size and (\ref{eq:IntDiff}) in the thermodynamic limit. 

\ni In order to clarify this point let us introduce some standard notations. We consider our system at initial time $t_0$ in equilibrium at inverse temperature $\b$. The partition function reads
$$
Z_N(\b)\equiv \Tr\left[e^{-\b H_N(t_0)}\right],
$$
and the thermal average of an operator $A$ is defined by
$$
\meanv{A}_{\b,t_0}\equiv \frac{1}{Z_N(\b)}\Tr\left[Ae^{-\b H_N(t_0)}\right].
$$
In particular, due to the fermion statistics, we have
\bea
\meanv{a^\dg_q a_p}_{\b,t_0}&=&\frac{\d_{qp}}{1+e^{\b \wp}}\nn\\
\meanv{a^\dg_q a^\dg_p}_{\b,t_0}&=&0\nn\\
\meanv{a_q a_p}_{\b,t_0}&=&0,\nn
\eea
where $\wp$ are the initial eigen-frequencies of the system. Since the Hamiltonian is quadratic we know that at least in principle it can be set in a diagonal form by a unitary transformation.

\ni Following the classical reference \cite{Ru69}, we introduce a generic time dependent observable as a function $f_N$ from a subset of the grid $S\subseteq\{1,..,N\}$ over the algebra of creation and annihilation operator. We allow this function to have also a possible explicit dependence on time, even if this will play no role in our exposition. Thus we write
$$
f_N(S; t)=\sum_{V\subseteq S}f(V; t)\frac{1}{N^{|V|}}\sum_{q_j,p_j,j\in V} :\prod_{j\in V} \eta_{q_j}\xi_{q_j}:e^{\sum_j i(p_j-q_j)j}+h.c.
$$ 
and $:\::$ denotes as usual the Wick product for the operators
\bea
\eta_q&=&a^\dg_q+a_{-q}\nn\\
\xi_q&=&a^\dg_{-q}-a_{q}\nn.
\eea

\ni For sake of brevity we will skip hereafter the locution $+ h.c.$ to indicate that every operator must be self adjoint. The crucial point here is that the Hamiltonian is quadratic in creation and annihilation operators. Thus the thermal average is evaluated by using the Wick theorem:
\bea
\meanv{f(S; t)}_{\b,t}&=&\sum_{V\subseteq S}f(V; t)\frac{1}{N^{|V|}}\sum_{q_j,p_j,j\in V} \meanv{:\prod_{j\in V} \eta_{q_j}\xi_{q_j}:}_{\b,t} e^{\sum_j i(p_j-q_j)j}+h.c.\nn\\
&=&\sum_{V\subseteq S}f(V; t)\frac{1}{N^{|V|}}\sum_{q_j,p_j,j\in V}e^{\sum_j i(p_j-q_j)j}\sum_{\mbox{all pairings }\Pi_V}(-1)^{\pi'}\prod_{(h,k)\in \Pi_V} \meanv{:\eta_{q_h}\xi_{q_k}:}_{\b,t}\nn
\eea
where $\pi'$ gives the parity of a given pairing $\Pi_V$ of the set $\{q_j,p_j:j\in V\}$. However we can take advantage by switching to the Heisenberg picture: we transfer the time dependence from the thermal average to the operators, and compute the thermal average at initial time $t_0$. Since time evolution is given by a Bogoliubov-Valatin semigroup, we have
\bea
\eta_q(t)&=&\sum_p \left(A^*_{qp}(t)+B_{-q,p}(t)\right)a^\dg_p+\left(B^*_{qp}(t)+A_{-q,p}(t)\right)a_{p}\nn\\
\xi_q(t)&=&\sum_p \left(A^*_{-q,p}(t)-B_{q,p}(t)\right)a^\dg_p+\left(B^*_{-q,p}(t)-A_{q,p}(t)\right)a_{p}\nn
\eea
and so
\be
\meanv{f(S; t)}_t=\sum_{V\subseteq S}f(V; t)\frac{1}{N^{|V|}}\sum_{q_j,p_j,j\in V}e^{\sum_j i(p_j-q_j)j}\sum_{\mbox{all pairings }\Pi_V}(-1)^{\pi'}\prod_{(h,k)\in \Pi_V} \meanv{:\eta_{q_h}(t)\xi_{q_k}(t):}_{\b,t_0}\nn
\ee

\ni For a given pair $\bar q, \bar p$ the only contributing term is
\be\label{eq:stato-1}
\sigma_{\bar q\bar p}(\b,t)=\frac1N\sum_{q'}\frac{A^*_{\bar qq'}(t)B^*_{q'-\bar p}(t)-B_{-\bar qq'}(t)A_{\bar q'p}(t)+B^*_{-\bar qq'}(t)B_{-\bar q'p}(t)-A^*_{\bar q q'}(t)A_{\bar q'p}(t)}{1+e^{\b\w_{q'}}}.
\ee
Therefore the time dependence of the state of the system is all in that formula. We have obtained
\be
\meanv{f(S; t)}_t=\sum_{V\in S}f(V; t)\Sigma(V,\b,t),
\ee
with
\be
\Sigma(V,\b,t)=\frac{1}{N^{|V|}}\sum_{q_j,p_j,j\in V}e^{\sum_j i(p_j-q_j)j}\sum_{\mbox{all pairings }\Pi_V}(-1)^{\pi'}\prod_{(h,k)\in \Pi_V} \sigma_{\bar q_h\bar p_k}(\b,t).
\ee
This is the time dependent state of the system. We notice that as long as $N$ remains finite, the dynamics is well defined and our derivation of the state is exact. The problem arises as $N\to\infty$: in this case, as we have seen in the previous Section, we need a further assumption on the interaction in order to have existence and uniqueness of the flow.

\ni To provide a simple example, we see how the number operator $\mathcal{N}=\frac1N\sum_j c^\dg_jc_j$ reads:
\bea
\meanv{\mathcal{N}}_t&=&\meanv{\mathcal{N}(t)}=\frac{1}{N^3}\sum_{j} \sum_{q,p}e^{ij(q-p)}\meanv{a^{\dg}_q(t) a_p(t)}\nn\\
&=&\int_0^{2\pi}\frac{dq}{1+e^{\b\wq}}\int_0^{2\pi} dp \left(|A(q,p,t)|^2-|B(q,p,t)|^2\right).\label{eq:mz-glob}
\eea

\ni As a final remark we observe that in the hypotheses of the global existence theorem, the state asymptotically approaches a limit, that is the stationary state of the system. But it is the nature of the time dependent Hamiltonian that determines whether it is an equilibrium state, viz. an ergodic theorem holds, or it is a non equilibrium stationary state. An example of this phenomenon is global versus local transverse perturbations in the XY chains, as it will be further discussed below.

\section{Preliminaries on Spin Chains}

\ni Now we turn our attention to quantum spin systems, ruled by nearest neighbour two body ferromagnetic interaction on $\mathbb{Z}$. We will be interested specifically in those models admitting a mapping into quasi free fermions analysed before. The most general Hamiltonian for describing such a class of systems is
\be
H_N=-\sum_{j=1}^{N^*} (J_j^x(t) S_j^xS_{j+1}^x+J_j^y(t) S_{j}^yS_{j+1}^y)-\sum_{j=1}^{N}h^z_j(t)S_j^z\label{eq:HXYZ-nonhom}.
\ee
The space this operator acts on is a tensor product of $1/2$-spin vector spaces $\mathcal{H}_j=\mathbb{C}^2$, each spanned by the vectors \textit{spin up} and \textit{spin down}: $\mathcal{H}_N\equiv\bigotimes_{j=1}^{N} \mathcal{H}_j={\mathbb{C}^2}^{\otimes N}$; the corresponding matrix algebra of $2\times 2$ matrices $GL_2(\mathbb{C})$ is spanned by the Pauli matrices $\sx,\sy,\sz$ plus the identity $\mathbb{I}$. As usual the thermodynamic limit for the system is  performed in the Fock space, defined as $\mathcal{F}\equiv\bigoplus_N \mathcal{H}_N$. The spin operators attached to each site $j$, $S_j^x$, $S_j^y$, and $S_j^z$, are defined in terms of Pauli matrices as $S_j^i=\frac{1}{2}\sigma_i$, $i=x,y,z$. The boundary conditions are defined by the value of $N^*$: \eg if $N^*=N-1$ we are dealing with open (or free) boundary conditions at extrema, while if $N^*=N$ we mean that $N+1=N$, so periodic boundary conditions are assumed. The following notations is usually adopted:
$
A_j\equiv \mathbb{I}\otimes\mathbb{I}\otimes...\otimes A\otimes \mathbb{I}\otimes ... \otimes\mathbb{I},
$ \ie the operator $A_j$ acts as $A$ on the Hilbert space of the $j$-th spin and as the identity on the others. So each observable, for finite $N$, belongs to the tensor product algebra $GL_2(\mathbb{C})^{\otimes N}$. Finally we will assume the same space and time regularity for the couplings as in Section 2.

\ni As mentioned above, these spin chains correspond to particular quadratic forms in the Fermi operators, the mapping being given by the Jordan-Wigner transformation, introduced in \cite{JW}. We put
$$
\left\{
\begin{array}{lll}
c_j&=&\frac{1}{2}(\sx_j-i\sy_j)\bigotimes_{k=1}^{j-1}(-\sz_k)\nonumber\\
c^\dg_j&=&\frac{1}{2}(\sx_j+i\sy_j)\bigotimes_{k=1}^{j-1}(-\sz_k)\nonumber,
\end{array}
\right.
$$
and their inverses
$$
\left\{
\begin{array}{lll}
\sx_j&=&(c_j^\dg+c_j)\bigotimes_{k=1}^{j-1}(-\sz_k)\\
\sy_j&=&-i(c_j^\dg-c_j)\bigotimes_{k=1}^{j-1}(-\sz_k)\\
\sz_j&=&2c^\dg_jc_j-\mathbb{I}\nn.
\end{array}
\right.
$$
It is easily seen that the $c$ operators satisfy fermionic anti-commutation relations:
\be
[c_j,c_k]_+=0, \qquad[c^\dg_j,c_k]_+=\d_{jk}.
\ee

\ni The morphism of algebras naturally induces a morphism of spaces: each $c_j$ ($c_j^\dg$) acts on a two level vector space $\mathcal{H}'_j$ as an annihilation (creation) operator, spanned by the vectors $\ket{0}_j$ (hole) and $\ket{1}_j$ (particle), as usual in the theory of Fermi systems. It is important to notice that this map does not preserve locality.

\ni Therefore, up to a constant term, the Hamiltonian (\ref{eq:HXYZ-nonhom}) is transformed in the following quadratic Hamiltonian for a one dimensional Fermi gas:
\be\label{eq:H-inhom-free}
H_N=-\sum_{j=1}^{N^*}(g+g_j^{xy}(t))(c^\dg_jc_{j+1}-c_jc^{\dg}_{j+1})-\sum_{j=1}^{N^*}\g_j^{xy}(t)(c^\dg_jc^\dg_{j+1}-c_jc_{j+1})-\sum_{i=1}^N (h+h^z_j(t))c^{\dg}_jc_j.
\ee
Here we will make a particular choice: we will look at impurities as (time dependent) perturbations of the isotropic XY Hamiltonian in a transverse field. The reason why we opt for perturbing around the isotropic instead of the (somehow more general) anisotropic model is merely technical and it will be clear in the following. We put
\be
g+g_j=\frac{J^x_j+J^y_j}{2},\quad \g_j=\frac{J^x_j-J^y_j}{2},
\ee
and $\sum_j g_j=\sum_j \g_j=\sum_j h_j=0$. In other words, the impurities are thought as fluctuations around the XX Hamiltonian. However in our framework it is more convenient to work in the Fourier space. By setting
\be
\tilde g_{q-p}=\sum_{k=1}^{N^*} g_{k,k+1}e^{ik(q-p)};\:\:
\tilde \g_{q+p}=\sum_{k=1}^{N^*} \g_{k,k+1}e^{ik(q+p)};\:\:
\tilde h_{q-p}=\sum_{k=1}^{N^*} h_{k}e^{ik(q-p)}\nn,
\ee
the Hamiltonian (\ref{eq:H-inhom-free}) can be written as
\bea\label{eq:HSpin-q-space}
H_N&=&\frac{1}{N^2}\sum_{q,p\in\mathbb{Z}(N)}\left[(g\cos q+h)\d_{qp}+h_{q-p}(t)+e^{iq}g_{q-p}(t)(1+e^{-i(q-p)})\right]a_q^{\dg}a_p\nn\\
&+&\left[\g_{q+p}(t)\left(\frac{e^{iq}-e^{-ip}}{2}\right)\right]a_q^{\dg}a^{\dg}_p+h.c.\nn
\eea

\section{Integral Equations for the Dynamics of Spin Chains}

\ni For spin chains we can identify 
\bea
\a_{qp}&=&(g\cos q+h)\d_{qp}+h_{q-p}(t)+e^{iq}g_{q-p}(t)(1+e^{-i(q-p)}),\\
\b^*_{qp}&=&\g_{q+p}(t)\left(\frac{e^{iq}-e^{-ip}}{2}\right).
\eea

\ni We would like to apply the theory developed in Section 2 to the last class of Hamiltonians. Rather then the dynamical variables $A_{qp}(t), B_{qp}(t)$, it turns out more convenient to deal with the site variables
\be
X_{k,q}(t)=\frac1N\sum_{q'}A_{qq'}(t)e^{-ik(q-q')},\quad V_{k,q}(t)=\frac1N\sum_{q'}B_{qq'}(t)e^{-ik(q+q')}.
\ee
These are in fact conceived to take into account the perturbation in the dynamics due to the impurities. The systems of equations (\ref{eq:BVflux2}) or (\ref{eq:IntDiff}) can be written in terms of these new variables as
\bea
i\dot A&=& \Hom(A,B)+\Inhom(X,V)\nn\\
i\dot B&=& \Hom(A,B)+\Inhom(X,V)\nn
\eea
where with $\Hom$ (respectively $\Inhom$) we have denoted the part of (\ref{eq:BVflux2}), (\ref{eq:IntDiff}) depending on homogeneous (inhomogeneous) couplings. Here it is important that $X_k(t), V_k(t)$ enter only in the inhomogeneous part of the last expressions. By Duhamel formula we obtain $A_{qp}(t)$ and $B_{qp}(t)$ in an integral form
\bea
A_{qp}(t)&=&e^{-i\wp(t-t_0)}\d_{qp}-\frac{i}{N}\sum_{k}e^{-i(q-p)k}\int_{t_0}^t dt' e^{-i\wp(t-t')}\Big[h_k(t') X'_{k,q}(t')\nn\\
&+&g_k(t')(e^{-iq}X'^{k+1}_q(t')+e^{ip}X'_{k,q}(t'))+\g_k(t')(e^{-iq}V'^{k+1}_q(t')-e^{ip}V'^k_q(t'))\Big]\nn\\
B_{qp}(t)&=&-ie^{-i\wp(t-t_0)}\d_{q,-p}+\frac{i}{N}\sum_{k}e^{-i(q+p)k}\int_{t_0}^t dt' e^{-i\wp(t-t')} \Big[h_k(t')V'^k_q(t')\nn\\
&+&g_k(t')(e^{-iq}V'^{k+1}_q(t')+e^{ip}V'^k_{q}(t'))+\g_k(t')(e^{-iq}X'^{k+1}_q(t')-e^{ip}X'_{k,q}(t'))\Big]\nn.
\eea
Here one can see the structure of the interaction. The quantities $g_k(t), \g_k(t)$ are linked respectively to the variable $X,V$ of the sites $k,k+1$, which reflects the nearest neighbour two body inhomogeneous interaction. On the other hand the transverse field is associated to the each single site. The eigen-frequencies of the related homogeneous system $\wp$ depend only on $g,h$. Of course we could allow $g,h$ to depend on time as well, simply by substituting every $\wp(t-t_0)$ with $\int_{t_0}^{t}dt'\wp(t')$. Nevertheless this would just make the theory more complicated, without adding any substantial new feature. Therefore we will suppose that $g,h$ are fixed. With a further step we can obtain a closed system of integral equations for the site variables, simply by using the definition of $X_k(t)$ and $V_k(t)$. We will skip all the calculations and give only the result in the limit $N\to\infty$:
\bea
X_k(q; t)&=&e^{-i\wq(t-t_0)}-i\sum_{j}\int_{t_0}^t dt'e^{iq(k-j)-ih(t-t')}\Big[e^{i\phi_{k-j-1}}J_{k-j-1}(t-t')\left(g_j(t')X^j(q; t')-\g_j(t')V^j(q; t')\right)\nn\\
&+&e^{i\phi_{k-j}}J_{k-j}(t-t')\left(h_j(t')X^j(q; t')+e^{-iq}g_j(t')X^{j+1}(q; t')+e^{-iq}\g_j V^{j+1}(q; t')\right)\Big]\label{eq: XkVk}\\
V^k(q; t)&=&-ie^{-i\wq(t-t_0)}+i\sum_{j}\int_{t_0}^t dt'e^{-iq(k-j)-ih(t-t')}\Big[e^{-i\phi_{k-j-1}}J^*_{k-j-1}(t-t')\left( g_j(t')V^{j}(q; t')-\g_j(t')X^j(q; t')\right)\nn\\
&+&e^{-i\phi_{k-j}}J^*_{j-k}(t-t') \left(h_j(t')V^j(q; t')+e^{-iq}g_j(t')V^{j+1}(q; t')+e^{-iq}\g_j X^{j+1}(q; t')\right)\Big].\label{eq: VkXk}
\eea
\ni Here we have used (see for instance \cite{TOI})
$$
\frac{1}{2\pi}\int_{0}^{2\pi} dp e^{-ipn} e^{-i(g\cos p+h)(t-t')}= e^{-ih(t-t')+i\phi_n}J_n(g(t-t')),
$$
where $J_n(t)$ are the Bessel functions of first kind and order $n$ (see \cite{AS64} for definitions and properties), which appear quite naturally in this context, and $\phi_n=n\pi/2$. It is remarkable that we can profit from this exact representation only by adopting the assumption $\g=0$: in this way the unperturbed spectrum is proportional to $\cos q$; otherwise it assumes a more involved form and we can no longer identify a known special function. This essentially concludes our analysis. In the rest of this Section we will point out some further mathematical detail in the theory. 

\ni The variables $X,V$ are the marginal Fourier transform of the former $A,B$ with respect to one argument. In the thermodynamic limit they are $\ell_{2,k}L_{2,q}C_t(\mathbb{Z}\times[0,2\pi]\times[t_0, T])$ functions: this is the natural space where seeking a solution. In addition it is worthwhile to notice that one can recover the $A,B$ variables via a simple Fourier transform in order to obtain the formula for the state of the system as in (\ref{eq:stato-1}).

\ni The main feature of our characterisation of the dynamics is to capture the properties of the chain right via the presence of the Bessel functions. They witness indeed the internal structure of the chain that matches with the external perturbation given by the impurities. So we have a very interesting effect, that can be described as follows: ergodicity is easily broken by the dynamics as one can see by the non thermalisation of observables, as the local transverse magnetisation (see for instance \cite{ABGM} or the Appendix \ref{app:B}); and indeed there is a strong memory effect given by the time-convolution integral in (\ref{eq: XkVk}) and (\ref{eq: VkXk}) with the Bessel functions $J_n(t)$, which naturally gives the time irreversibility of the dynamics. 

\ni Finally it is useful to recall the following formula (from formulas 6.671 in \cite{TOI}):
\be\label{eq:f}
\hat J_n(\Omega)\equiv\int_{0}^{+\infty} d\t J_n\left(g\t\right)e^{i\Omega\t}=\frac{\chi(|\Omega|\leq g)e^{in\arcsin(\Omega/g)}+ig^ne^{i\phi_n}\chi(|\Omega|\geq g)/(\sqrt{\Omega^2-g^2}+\Omega)^n}{\sqrt{|\Omega^2-\left(\frac g2\right)^2|}},
\ee
that permits us to represent
$$
J(t)=\int_{0}^{+\infty}d\Omega e^{-i\Omega t}\hat J_n(\Omega).
$$
Thus we can decompose the perturbations in frequencies (via Fourier transform) and study the problem of resonances between the three quantities in the play: the proper frequencies of the perturbation, the strength of the perturbation, the proper frequencies of the chain. Such a scheme is particularly evident in the instance of perturbations periodic or quasi periodic in $t$. This is just a conceptual picture and to implement it in a concrete example is quite a hard task, even in the simplest cases. We will give a hint of that in the sequel. Nevertheless we believe that it is rather explicative: in general of course the inverse of a continuous function via Fourier transform is not integrable; in our framework this case may arise only because of resonances, manifesting themselves by means of non integrable singularities. Existence of the dynamics will correspond to a proper renormalisation of these resonances. 

\section{Inhomogeneous Transverse Field in the XY Chain}

\ni In this last Section we will focus on the explicit form the system of equations (\ref{eq: XkVk}), (\ref{eq: VkXk}) takes when the impurities act only at the level of the transverse field. Our treatment will be twofold: at first we will deal with the XY chain, giving a more exhaustive account of the choice $\g=0$; then we will turn to the dynamics of the XX chain, that has a considerably simpler form, nonetheless capturing all the main features (encoded in the Bessel functions). We will show explicitly the mechanism behind existence of local and global solutions by briefly discussing the case of oscillation of a single impurity. 

\ni For the XY model with impurities in the transverse field we have
\be\label{eq: XY-imp-alpha,beta}
\a_{qp}=g(\cos q+h)\d_{qp}+h_{q-p}(t),\quad \b_{qp}=-i\g\sin q\d_{p,-q}.
\ee
Therefore the evolution equation for the dynamics reads
\bea
i\dot A_{qp}(t)&=&A_{qp}(t)(g\cos p+h)+i\g\sin p B_{q,-p}(t)+\frac1N\sum_{q'}A_{qq'}(t)h_{q'-p}(t)\label{eq: XX-A}\\
i\dot B_{qp}(t)&=&-B_{qp}(t)(g\cos p+h)-2i\g\sin p A_{q,-p}(t)-\frac1N\sum_{q'}B_{qq'}(t)h_{q'-p}(t)\label{eq: XX-B}.
\eea

\ni It is convenient to analyse in primis the unperturbed system ($h=0$) and reduce the system in diagonal form. That is done via the same transformation one introduces in order to diagonalise the Hamiltonian (this procedure appeared for the first time in the paper of Lieb, Mattis and Schultz \cite{LMS} and it is reviewed in Appendix \ref{app-A}). It is more practical to arrange (\ref{eq: XX-A}) and (\ref{eq: XX-B}) in the vectorial form
$$
i\dot{\mathcal{A}}_{qp}(t)=\Gamma_p \mathcal{A}_{qp}(t_0),\quad\mbox{with}\quad\mathcal{A}_{qp}= \left(
\begin{array}{c}
 A_{q,p}\\
 B_{q,p}\\
 A_{q,-p}\\
 B_{q,-p}
\end{array}
\right)
$$
and
$$
\Gamma_p=\left(\begin{array}{rrrr}
(g\cos p+h)& 0 & 0 & -i\g\sin p\\ 
0 & -(g\cos p+h) & -i\g\sin p & 0\\
0 & i\g\sin p & (g\cos p+h) & 0\\
i\g\sin p & 0 & 0 & -(g\cos p+h)
\end{array}\right).
$$
This is in fact the matrix to be diagonalised in order to find the spectrum of the model. This is done by means of the unitary matrix (\ref{eq:W_q}), and the eigenvalues for the energy levels are given by (\ref{eq:E^2_q}). Thus the system reads
$$
i\dot{\mathcal{A}}'_{qp}(t)=\left(\begin{array}{rrrr}
E_p& 0 & 0 & 0\\ 
0 & -E_p & 0 & 0\\
0 & 0 & E_p & 0\\
0 & 0 & 0 & -E_p
\end{array}\right)\mathcal{A}'_{qp}(t_0)
$$
solved by
\bea
A'_{qp}&=&e^{-iE_p(t-t_0)}A'_{qp}(t_0),\quad A'_{qp}(t_0)=\d_{qp}\cos\phi_q,\nn\\
B'_{qp}&=&e^{iE_p(t-t_0)}B'_{qp}(t_0),\quad B'_{qp}(t_0)=-i\d_{q,-p}\sin\phi_q.\nn
\eea
In the original variables the solution is instead given by
\bea
A_{qp}(t)&=&\d_{qp}\left(\cos(E_p(t-t_0))-i\frac{g\cos p+h}{E_p}\sin(E_p(t-t_0))\right)\nn\\
B_{qp}(t)&=&-\g\frac{\sin p}{E_p}\d_{q,-p}\sin(E_p(t-t_0)).\nn
\eea
We note that, by plugging the last two expression in (\ref{eq:mz-glob}), we easily recover the non-ergodicity of the transverse magnetisation $m^z$ firstly achieved in \cite{BMD} (see also \cite{Leb}\cite{Ig}). We have
\bea
\int dp|A_{qp}(t)|^2&=&\cos^2(E_q(t-t_0))+\frac{(g\cos q+h)^2}{E^2_q}\sin^2(E_p(t-t_0))+\nn\\
\int dp|B_{qp}(t)|^2&=&\left(\g\frac{\sin q}{E_{q}}\sin(E_{q}(t-t_0))\right)^2,\nn
\eea
and so
\be\label{eq:mz-XYnon-th}
1+\meanv{m^z}=\int\frac{dq}{1+e^{\b E_q}}\cos^2(E_q(t-t_0))\left[1+\left(\frac{(g\cos q+h)^2}{E^2_q}-\g^2\frac{\sin^2q}{E^2_{q}}\right)\tan^2(E_q(t-t_0))\right]
\ee
We notice two important features of this formula: because of the $-$ sign in (\ref{eq:mz-glob}), the limit $h\to0$ does not give the equilibrium value $m^z$; in the isotropic limit $\g\to0$ one recovers the thermalisation of $m^z$, which mirrors $[H_{XX}, m^z]=0$, the conservation of the magnetic momentum in the XX chain. As for the longitudinal magnetisation the situation is more involved, due to its more complicate representation in terms of Fermi operators. However it is remarkable that this observable approaches its equilibrium value for $t\to\infty$ \cite{B2}\cite{Ig}.

\ni When a perturbative transverse field is added, we describe the system in terms of $X,V$ variables. Since the relations linking $A,B$ with $X,V$ are linear, we have
\be
X'_{k,q}(t)=\sum_{q'}A'_{qq'}(t)e^{-ik(q'-q)},\quad V'_{k,q}(t)=\sum_{q'}B'_{qq'}(t)e^{-ik(q'+q)},
\ee
and also the following equations for the eigenvariables
\bea
i\dot A'_{qp}(t)&=&E_pA'_{qp}(t)+\frac{1}{N}\sum_{k}h_k(t)X'_{k,q}(t)e^{-ik(q-p)}\label{eq:Eq_tXYA'}\\
i\dot B_{qp}(t)&=&-E_pB'_{qp}(t)-\frac{1}{N}\sum_{k}h_k(t)V'_{k,q}(t)e^{-ik(q-p)}\label{eq:Eq_tXYB'}.
\eea
Now these equations are decoupled and they can be solved separately by Duhamel principle:
\bea
A'_{qp}(t)&=&e^{-iE_p(t-t_0)}\d_{qp}\cos\phi_q-\frac{i}{N}\sum_{k}e^{-i(q-p)k}\int_{t_0}^t dt' e^{-iE_p(t-t')} h_k(t')X'_{k,q}(t')\label{eq:sol_tXYA'}\\
B'_{qp}(t)&=&-ie^{-iE_p(t-t_0)}\d_{q,-p}\sin\phi_q+\frac{i}{N}\sum_{k}e^{-i(q+p)k}\int_{t_0}^t dt' e^{-iE_p(t-t')} h_k(t')V'^k_q(t')\label{eq:sol_tXYB'}.
\eea
As we have already seen (here $\wp=E_p$), there are closed equations for $X'_{k,q},V'^k_q$:
\be
X'_{k,q}(t)=e^{-i\wq(t-t_0)}\cos\phi_q-\frac{i}{N}\sum_{j}\int_{t_0}^t dt'\left(\sum_p e^{i(q-p)(k-j)} e^{-i\wp(t-t')}\right) h_j(t')X'^j_q(t')
\ee
and
\be
V'^k_q(t)=-ie^{-i\wq(t-t_0)}\sin\phi_q+\frac{i}{N}\sum_{j}\int_{t_0}^t dt'\left(\sum_p e^{i(q+p)(k-j)} e^{-i\wp(t-t')}\right) h_j(t')V'^j_q(t').
\ee
These equations, although formally identical to the ones we got in the previous Section, are technically more involved, essentially because
$$
\lim_N\frac{1}{N}\sum_p e^{i(q\pm p)(k-j)} e^{-iE_p(t-t')}=\frac{1}{2\pi}\int_{-\pi}^{\pi}dp e^{i(q\pm p)(k-j)} e^{-iE_p(t-t')},
$$
due to the non linear dependance of $E_p$ on $\cos p$, is not a known function. Thus for the XY chain (\ie for $\g\neq0$) we still have a clear theoretical picture for the dynamics, but in general we do not have an exact analytical formulation of it.

\vspace{0.5cm}

\ni In the XX Hamiltonian there is a further simplification: if we set $\wq=(g\cos q+h)$, we have
\be
\a_{qp}(t)=\wq\d_{qp}+h_{q-p}(t),\quad\b_{qp}=0.
\ee

\ni Since the term $\b_{qp}$ vanishes, equations (\ref{eq:BVflux2}), (\ref{eq:IntDiff}) decouple (they are in fact the same equation), and the dynamics of the system is described by $A_{qp}(t)$. Hence the equation we are interested in is
\be\label{eq:eqEvXX1}
i\dot A_{qp}(t)=A_{qp}\wq+\frac1N\sum_{q'}A_{qq'}(t)h_{q'-p}(t).
\ee
In terms of the site variables previously introduced, this translates into a system of equations involving only the $X_{k,q}(t)$ associated to the $k$-th impurity. At finite $N$ we get
\be\label{eq:eqEvXX-SolF}
X_{k,q}(t)=e^{-i\wq(t-t_0)}-\frac{i}{N}\sum_{j}\int_{t_0}^t dt'\left(\sum_p e^{i(q-p)(k-j)} e^{-i(g\cos p+h)(t-t')}\right) h_j(t')X^j_q(t').
\ee
Passing to the limit $N\to\infty$ equations (\ref{eq: XkVk}), (\ref{eq: VkXk}) simplify a lot 
\be\label{eq:eqEvXX-SolFmany}
X_k(q,t)=e^{-i\wq(t-t_0)}- i\sum_{j} e^{iq(k-j)} \int_{t_0}^t dt' e^{-ih(t-t')}J_{j-k}(g(t-t')) h_j(t')X^j(q,t').
\ee
The simplest, but already quite rich, case to study in this context is the one of a single impurity localised in the site $\hat k$,$h_j(t)=\d_{j\hat k}h(t)$, for which
\be\label{eq:eqEvXX-1imp-k}
X_k(q;t)=e^{-i\wq(t-t_0)}-ie^{iq(k-\hat k)}\int_{t_0}^t dt' e^{-ih(t-t')}J_{k-\hat k}(g(t-t')) h(t')X^{\hat k}(q; t').
\ee
It is immediate to see that the state of the system is completely determined by $X^{\hat k}(q; t)\equiv X(q; t)$. This is defined by
\be\label{eq:eqEvXX-1imp-UNo}
X(q;t)=e^{-i\wq(t-t_0)}-i\int_{t_0}^t dt' J_{0}(g(t-t')) h(t')X(q; t').
\ee
Here we have incorporated in the field its zero mode, and so our unperturbed system has now eigen frequencies $\wq=g\cos q$, a choice somehow natural in this simplified situation. 

\ni We can now differentiate this integral equation in order to obtain the linear Schr\"odinger equation
\be\label{eq:schr}
i\partial_t X(q;t)=(\wq+h(t)) X(q;t)-g\int_{t_0}^{t} d\t \left(gJ_1\left(g(t-\t)\right)-i\wq J_0\left(g(t-\t)\right)\right)h(\t)X(q;\t),
\ee
which reads more familiarly in the configuration space of the chain $\mathbb{Z}$
\be\label{eq:schrII}
i\partial_t X_k(t)=(-g\D+h(t)) X_k(t)-W[X,\D X; t].
\ee
Here $-\D$ is the Laplace operator on $\mathbb{Z}$, with eigenvalues $\cos q$, and 
$$
W[X,\D X; t]=g\int_{t_0}^{t} d\t \left(gJ_1\left(g(t-\t)\right)-i J_0\left(g(t-\t)\right)\D\right)h(\t)X_k(\t).
$$

\ni The equation (\ref{eq:schrII}) describes the evolution of a single spin coupled to an external field $h(t)$ and to the rest of the chain via the parameter $g$. It is a Schr\"odinger equation with Floquet Hamiltonian, but the time dependance exhibits explicitly a gentle memory effect via the convolution operator. This eventually makes it easier to study with respect to the case of the purely multiplicative potential. 
In its integral form it can be symbolically rewritten as $X=X_0+KX$, where $K$ is an integral operator acting on continuous functions (in $t$). Equations of this form are usually approached by seeking a solution like $X=(1-K)^{-1}X_0$ by means of Fredohlm theory (for an exhaustive treatment we refer to the original paper by I. Fredholm and the monograph by I. Gohberg, S. Goldberg and M. A. Kaashoek \cite{int}). Now we switch on the field, for example continuously, at time $t_0$, and let the system evolve up to time $T>t_0$. The integral operator has a continuous kernel in each closed interval $[t_0,T]$: this automatically implies existence of the solution and we also have the explicit form of it by Fredholm expansion. This is morally the local existence theorem of Section 2. But as we send $T\to\infty$ the situation becomes different. Continuity and boundedness of the perturbation are no longer sufficient and necessary conditions in order to ensure existence, and one needs for example a suitable relaxation property at infinity. A possibility is, as already discussed, that the limit $\lim_{t\to\infty} h(t)=\bar h$ exists. In this case the results reported in \cite{ABGM} (see also \cite{Bar}) show that the existence of the solution is not affected by the specific time dependence of the perturbation: the system thermalise to an equilibrium state given by the limiting Hamiltonian. It is remarkable that the situation is very different for a perturbation of the same time dependance, but spatially homogenous on the whole chain. In this occurrence the system thermalise as well, but the final state is not an equilibrium state, meaning that the transverse magnetisation does not go to zero in the limit $h\to0$, once the asymptotic limit $T\to\infty$ has been taken. As mentioned in the Introduction, this aspect is exhaustively reviewed in \cite{Bar}\cite{Leb} it is discussed in relation to more general dynamical properties of simple classical and quantum systems.

\ni More challenging and still unsolved is instead the case of transverse field with an asymptotic oscillating behaviour. Even the occurrence (somehow basic in this context) of time periodic perturbation is hard to manage, although partial results have been achieved: \cite{ABGM}, \cite{CG}. These situations indeed do not fall in the hypotheses required by the global existence theorem stated in Section 2. Therefore a proof of the existence of the dynamics for all time will need some new ideas and additional tools. 

\ni However this problem is inscribed in a well known and vastly ranging topic in theoretical physics, namely semiclassical interaction between light and matter. In our framework we figure out the non autonomous spin systems as feeling the presence of an oscillating classical external field. As we have discussed, a rigorous mathematical approach to this question is hard to pursue and we can point out at least two (probably not independent) hard aspects of the problem: the global existence of the time evolution is ultimately related to the preservation of the infinite dimensional analogous of the KAM tori in presence of a perturbation; the analysis of the final state has possibly to take into account an infinite number of crossing of the critical point of the model.

\section{Outlooks}

\ni In this paper we have made an attempt to describe with a unified formalism the quantum motion of those spin chain models which can be mapped into free fermions, when the Hamiltonian explicitly depends on time. We have discussed on a general level the problem of existence of dynamics locally and globally in time. Then we have deepened the subject by studying the time evolution for spin chains with a more specific system of integral equations. 

\ni The central feature emerging by our analysis is that we can a priori distinguish two cases: on one hand we have time dependent perturbations approaching a determined limit in the extrema of the definition interval in time; in this scenario one has global existence of time evolution and the system approaches a final state. Therefore the main investigations are focused on the properties of such a state, which are far to be trivial in general (they are eventually connected to the quantum phase transition of the model). On the other hand we have the case of temporal dependance with asymptotic oscillations. We can single out this last situation as the more challenging one to deal with, because the possible global existence of the dynamics requires a more delicate study. 

\ni Of course in the context of spin chains the next natural step will be to obtain a similar picture in the case of the XYZ model, which is mapped into fermions with a quartic interaction. This class of models is certainly more appropriate to describe accurately the properties of low dimensional quantum systems and a fortiori the effect of the defects. One of the main obstacles to extend our analysis to these models is that it is not possible to recover the state so easily as we have done in Section 3, once one has the dynamical variables, by means of the Wick theorem. Therefore one should recur to a perturbative expansion which will render much more involved the whole analysis. However, in principle it is possible to suitably modify our approach to obtain (at least in a perturbative scheme) the equations of the dynamics in the case of fermions with quartic interaction. Then one could study the effects given by the presence of impurities, at least in some simplified case.

\vspace{1cm}

\ni {\bf Acknowledgements\\}
I would like to thank H. Narnhofer, for a useful correspondence about the papers \cite{narn}. Moreover I am grateful to A. Giuliani, F. Illuminati and especially to B. Schlein for many valuable comments and suggestions. 

\ni This work has been supported by the ERC Grant MAQD 240518.

\appendix

\section{Diagonalisation of the XY Chain}\label{app-A}

\ni This appendix is devoted to review the procedure of diagonalisation developed in \cite{LMS} (see also \cite{Bar}).  After a Fourier transform of the Fermi operators, the Hamiltonian of the XY model becomes (up to the irrelevant addendum $\sum_q e^{iq}=1$)
\bea
2H_N(\g,h)&=&-\sum_{q}(2g\cos q+2h)a^\dg_qa_q+e^{-iq}a^\dg_qa^\dg_{-q}-e^{iq}a_qa_{-q}\nn\\
&=&-\sum_{q}\left[(g\cos q+h)(a^\dg_qa_q+a^\dg_{-q}a_{-q})-i\g\sin q(a^\dg_qa^\dg_{-q}+a_qa_{-q})\right]\nn\\
&=&-\sum_q H_q(g,\g,h),
\eea
bearing in mind that
\bea
\sum_q g(q) \alpha_q\alpha_{-q}&=&\frac{1}{2}\sum_q (g(q)-g(-q))\alpha_q\alpha_{-q}\nn\\
\sum_q g(q) \alpha_q\beta_{q}&=&\frac{1}{2}\sum_q g(q)\alpha_q\beta_q+g(-q)\alpha_{-q}\beta_{-q}\nn\\
\alpha_q\beta_q&=&\frac{1}{2}[\alpha_q,\beta_q]+\frac{\delta_{qp}\delta_{\alpha\beta^{\dg}}}{2}\nn,
\eea
for every couple of Fermi operators such that $[\alpha_q, \beta_p]_+=\delta_{qp}$. We notice that all the $H_q$ acts in independent subspaces, that is 
$$
2H_q=\mathbb{I}\otimes...\otimes \left[(g\cos q+h)(a^\dg_qa_q+a^\dg_{-q}a_{-q})-i\g\sin q(a^\dg_qa^\dg_{-q}+a_qa_{-q})\right] \otimes...\otimes\mathbb{I}
$$
and so we always have $[H_p,H_q]=0$ $\forall p,q\in\mathbb{Z}(N)$. This simply means that we can diagonalise all the modes independently, thus we can operate by fixing $q$. Each $H_q$ is a quadratic form in creation and annihilation operators. In order to render it more symmetric, we will write (up to a constant term $\cos q+h$)
$$
4H_q=\frac{1}{2}\Big((g\cos q+h)([a^\dg_q,a_q]+[a^\dg_{-q},a_{-q}])-i\g\sin q([a^\dg_q,a^\dg_{-q}]+[a_q,a_{-q}])\Big),
$$
or in matrix form
\be\label{eq:HXYHom-Matr}
4H_q=\left(\begin{array}{rrrr}a^\dg_q& a_q& a^\dg_{-q} & a_{-q}\end{array}\right) \left(\begin{array}{rrrr}
\G_1& 0 & 0 & -i\G_2\\ 
0 & -\G_1 & -i\G_2 & 0\\
0 & i\G_2 & \G_1 & 0\\
i\G_2 & 0 & 0 & -\G_1
\end{array}\right) \left(
\begin{array}{c}
 a_q \\
 a^\dg_q  \\
 a_{-q}\\
 a^\dg_{-q}
\end{array}
\right),
\ee
with $\G_1=g\cos q+h$ and $\G_2=\g\sin q$. This matrix can be easily diagonalised, thereby obtaining the eigenvalues for the energy of the system:
\be\label{eq:E^2_q}
E^2_q=(g\cos q+h)^2+\g^2\sin^2 q.
\ee

\ni The Bogoliubov-Valatin transformation $W$ that diagonalises the matrix can be directly verified to be the tensor product of (SU(2)$\times$SU(2)) rotations around the $y$-axis:
\bea
W&=&\bigotimes_{q=-\pi}^{\pi}W_q,\nn\\
W_q&=&\left(\begin{array}{rr}
\mathbb{I}\cos\phi_q & -\sy\sin\phi_q \\ 
\sy\sin \phi_q &\mathbb{I}\cos\phi_q   
\end{array}\right)\label{eq:W_q},
\eea
that is the Hamiltonian is diagonal in the new operators
\bea
b_q&=&a_q\cos\phi_q+ia^\dg_{-q}\sin\phi_q\nn\\
b^{\dg}_q&=&a^\dg_q\cos \phi_q-ia_{-q}\sin \phi_q,
\eea
with $\cos 2\phi_q=(g\cos q+h)/E_q$, $\sin \phi_q=-\g\sin q/E_q$. Sometimes it can be useful to  get rid of imaginary unit and make everything real. Let us introduce the unitary matrix
$$
T=\left(\begin{array}{rr}
e^{i\frac{\pi}{4}} & 0 \\ 
0 & e^{-i\frac{\pi}{4}}  
\end{array}\right),
$$
with $\sy T=T^\dg\sy$, and $T^\dg\sy T=(T^\dg)^2\sy=-i\sz\sy=-\sx$, and define
$$
\hat  W_q=\left(\begin{array}{rr}
\mathbb{I}\cos\phi_q & -T^\dg\sy T\sin\phi_q \\ 
T^\dg\sy T\sin \phi_q &\mathbb{I}\cos\phi_q   
\end{array}\right)\nn,
$$
as the transformation that diagonalises our bilinear form with phase adjustment. Thus for the new operator obtained in this way we have
\bea
\hat  b_q&=&a_q\cos\phi_q+a^\dg_{-q}\sin\phi_q\nn\\
\hat  b^{\dg}_q&=&a^\dg_q\cos \phi_q+a_{-q}\sin \phi_q,
\eea

\ni The Hamiltonian in the new variables becomes
\bea
-\sum_q H_q&=&-\sum_q\frac{1}{4}\left(E_q [b^\dg_q,b_q]+E_q[b^\dg_{-q},b_{-q}]\right)\nn\\
&=&-\frac{1}{2}\sum_q\left(E_q b^\dg_q b_q+E_q b^\dg_{-q}b_{-q}\right)\nn\\
&=&-\sum_q E_q b^\dg_q b_q.
\eea
Analogously we find
$$
-\sum_q H_q=-\sum_q E_q \hat  b^\dg_q\hat  b_q.
$$
\ni It turns useful to specify the transformation that diagonalises our initial Fermi Hamiltonian in $c,c^\dg$: denoting by $b\equiv(b_q, b^\dg_q)$ and $c\equiv(c_j, c^\dg_j)$ one has 
$$
b=U^\dg  c U
$$
where $U=W\circ FT$ is the composition of a Fourier transform and our Bogoliubov-Valatin transformation. Thus the relations defining the matrix elements $U_{qj}$ are
\be\label{eq:U-inutile}
\left\{
\begin{array}{rrr}
b_q&=&\frac{1}{\sqrt{N}}\sum_{j=1}^N e^{iqj}\left(\cos\phi_qc_j+i\sin\phi_q c^{\dg}_j\right)\\
b_q^\dg&=&\frac{1}{\sqrt{N}}\sum_{j=1}^N e^{-iqj}\left(\cos\phi_q c^\dg_j-i\sin\phi_q c_j\right),
\end{array}
\right.
\ee
with the obvious (more useful) inverses:
\be\label{eq:U-utile}
\left\{
\begin{array}{rrr}
c_j&=&\frac{1}{\sqrt{N}}\sum_{q} e^{-iqj}\cos\phi_q b_q+ie^{iqj}\sin\phi_q b^\dg_q, \\
c_j^\dg&=&\frac{1}{\sqrt{N}}\sum_{q} e^{iqj}\cos\phi_q b^\dg_q-ie^{-iqj}\sin\phi_q b_q.
\end{array}
\right.
\ee

\section{Quench of an Impurity in the XX Chain}\label{app:B}

\ni In this appendix we provide a very simple example which witnesses the break of ergodicity of the dynamics of the XX chain. We will consider an abrupt change of transverse magnetic field in $t=t_0$ acting only on one $k$-th spin of the chain:
\be
h(t)=\left\{
\begin{array}{ccc}
0,&\quad&t\leq t_0;\\
h,&\quad&t>t_0.
\end{array}
\right.
\ee
We are going to study the time behaviour of the local transverse magnetisation. Since we deal with the XX model, the dynamics is described by the sole $A$ variables (see formulas (\ref{eq:alfa-beta}), (\ref{eq:BVflux2}) and (\ref{eq:IntDiff})). Following \cite{ABGM} we have for the local magnetisation along $z$ in the $k$-th site
\bea
1+\meanv{m^z_k(t)}&=&\frac{1}{N^2}\sum_{q,p}e^{ik(q-p)}\meanv{a_q^\dg(t)a_p(t)}\nn\\
&=&\frac{1}{N^4}\sum_{q,p}e^{ik(q-p)}\sum_{q',p'}A^*_{qq'}(t)A_{pp'}(t)\meanv{a_{q'}^\dg a_{p'}}\nn\\
&=&\frac{1}{N^3}\sum_{q'}\sum_{q,p}e^{ik(q-p)}\frac{A^*_{qq'}(t)A_{pq'}(t)}{1+e^{\b g\w_{q'}}},
\eea
where we recall $\wq=g\cos q$. Noticing that
$$
|X_{k,q}|^2=\frac{1}{N^2}\sum_{qp}e^{ik(q-p)}A^*_{qq'}(t)A_{pq'}(t),
$$
we immediately get as $N\to\infty$
\be\label{eq:m}
1+\meanv{m_k^z(t)}=\frac{2}{\pi}\int_{-\pi}^\pi \frac{dp}{1+e^{\b\wp}} |X_{k,p}|^2.
\ee
The variable $X(q; t)$ is the unknown of the equation (\ref{eq:schrII}). It is convenient to study it into its integral form:
\be\label{eq:X1}
X(q;t)=e^{-i\wq(t-t_0)}-ih\int_{t_0}^t dt' J_{0}(g(t-t')) X(q; t').
\ee
Let us perform the change of variables $Y(p;t)=X(p,t)e^{\wp(t-t_0)}$. Of course the magnetisation is unaffected by it, since it rests proportional to $|Y|^2$. The integral equation becomes
\be
Y_{t_0}(p; t)=1-ih\int_{0}^{t-t_0} d t' J_0(t')e^{i\wp t'}Y_{t_0}(p; t-t').
\ee
Now we would like to send $t_0\to-\infty$, that corresponds to study the asymptotic dynamics. The limiting equation reads
\be
Y(p; t)=1-ih\int_{0}^{+\infty} d t' J_0(t')e^{i\wp t'}Y(p; t-t').
\ee
Now we pass to Fourier transform in time, and we get
$$
(1+ih\hat J (\wp+\xi))\hat Y(\xi,t)=\d(\xi),
$$
where (see (\ref{eq:f}) for $n=0$)
$$
\hat J (\wp+\xi)=\frac{\chi(g<\wp+\xi)+i\chi(g>\wp+\xi)}{\sqrt{|g^2-(\wp+\xi)^2|}}.
$$
Then we obtain
\bea
Y(p,t)&=&\int d\xi e^{i\xi t}\d(\xi)\frac{1-ih\hat J (\wp+\xi)}{1+h^2|J (\wp+\xi)|^2}\nn\\
&=&\frac{1-ih\hat J (\wp)}{1+h^2|J (\wp)|^2}
\eea
Therefore the asymptotic value of $Y$ is constant in time and
\be
|Y(p;t)|^2=|Y(p;\infty)|^2=(1+h^2|J (\wp)|^2)^{-1}=\frac{\sin ^2p}{\sin ^2p+h^2/g^2}.
\ee
Hence
\bea
1+\meanv{m_k^z(\infty)}&=&\frac{2}{\pi}\int_{-\pi}^\pi \frac{dp}{1+e^{\b\wp(h_0)}} \frac{\sin ^2p}{\sin ^2p+h^2/g^2}\nn\\
&=&1+\meanv{m_k^z}_{0}-\frac{h^2}{g^2} \frac{2}{\pi}\int_{-\pi}^\pi \frac{dp}{1+e^{\b\wp(h_0)}}\frac{1}{\sin ^2p+h^2/g^2},
\eea
where we denoted by $\meanv{m_k^z}_{0}=0$ the initial equilibrium value of the magnetisation. Therefore we have a nice explicit formula
\be\label{eq:explicit}
\meanv{m_k^z(\infty)}=-h^2 \frac{2}{\pi}\int_{-\pi}^\pi \frac{dp}{1+e^{\b\wp(h_0)}}\frac{1}{g^2\sin ^2p+h^2}.
\ee
We can see immediately that as $h\neq0$ the magnetisation does not approach its initial value. An analysis of the opposite situation is also interesting. Consider the transverse field to be abruptly switched off form $h$ to $0$. By the aid of (\ref{eq:explicit}) we can only argue that the magnetisation in this case should thermalise. In fact the calculations are slightly more involved, because the equilibrium system is given by the XX chain with one impurity, which is fairly non trivial. It has to be diagonalised in order to recover the eigen-frequencies $\wp$ and the initial state. The spectrum of this system presents a band $\wp=g\cos q$ and an isolated eigenvalue $\wp=\sqrt{g^2+h^2}$. These results are in \cite{ABGM} (see also \cite{CG}), where it has been proven the ergodic behaviour the transverse magnetisation.

\end{document}